\documentclass[]{aanew}
\usepackage{psfig,graphicx}
\usepackage[authoryear]{natbib}
\bibpunct{(}{)}{;}{a}{}{,}
\begin{document}

\title{Stellar tracers of the Cygnus Arm}
\subtitle{I. Spectroscopic study of bright photometric candidates
\thanks{Based on observations made at 
Observatoire de Haute Provence (CNRS), France}}

\author{I.~Negueruela\inst{1,2}
\and A. Marco\inst{2}}                   
                                                            
\institute{Observatoire de Strasbourg, 11 rue de l'Universit\'{e},
F67000 Strasbourg, France
\and
Dpto. de F\'{\i}sica, Ingenier\'{\i}a de Sistemas y Teor\'{\i}a de
la Se\~{n}al, Universidad de Alicante, Apdo. 99, E03080 Alicante, Spain
}

\mail{ignacio@dfists.ua.es}

\date{Received    / Accepted     }

\titlerunning{OB Stars in Outer Arm}

\abstract{
We present medium-resolution spectroscopy of a sample of stars in the
second Galactic quadrant selected from the literature because their
colours suggest that they are moderately-reddened early-type stars at
very large distances. From the derived spectral types and observed
colours, we calculate distances to all these objects. For a sizable
fraction of our sample, we find distances well in excess of what is
expected for Perseus Arm objects, even allowing for rather generous
errors. In the interval $l=150\degr-180\degr$, there is a large number
of objects with distances in excess of 4 kpc, which are likely tracing
the Outer or Cygnus Arm. In particular, we find that the association
Cam OB3 is placed on this Arm. Based on our results, the extent and
definition of the associations Cas OB4 and Aur OB2 need to be
reevaluated. 
}
\maketitle 


\section{Introduction}

Since they are necessarily young, OB stars -- in particular, those
clearly belonging to an association or illuminating an \ion{H}{ii}
region -- trace the areas of recent star formation. As star formation
happens preferentially in the spiral Galactic arms, it has been
generally considered that O and early-B type stars can be used as
spiral-structure tracers, at least when they are studied in relatively
large numbers \citep[cf.][]{rus03}. After all, the first conclusive
evidence for the spiral  
structure of the Milky Way was derived from the distribution of
OB associations in the Solar neighbourhood \citep{mor52}.

In spite of several earlier efforts \citep[e.g.,][]{iss70,mav73}, an
outline of the Galactic spiral structure based on optical tracers was
only achieved by \citet{gag}, based fundamentally on data from
\ion{H}{ii} regions and their exciting stars. Their sketch of Galactic
structure included four spiral arms, with the Sun being located in a
minor feature or spur, generally known as the Orion Spur. The
main spiral structure is determined by the Sagittarius-Carina Arm
(i.e., the first arm towards the Galactic Centre from the position of
the Sun or $-$I arm) and the Norma ($-$III or internal) Arm. The two other 
arms, Perseus (the first arm towards the outside from the position of
the Sun or $+$I arm) and Scutum-Crux ($-$II) were suspected to be less
important features. 

The picture of the Galaxy today is more elaborate,
including a bar and a central ring \citep[e.g.,][]{val02}, as well as
a warp in the outer regions \citep[e.g.,][]{das01}, but the
basic design still survives. Though \citet{fer01} conclude that a
two-arm model cannot be ruled out with existing observations, most
modern studies favour a four-armed Galaxy \citep{val02,rus03}.

\begin{table*}[ht]
\caption{Distant OB stars observed from the OHP 1.52-m telescope. The
first column indicates the volume and number in the Luminous Stars
catalogue. The fourth and fifth columns indicate the dates when the
two spectral positions were observed and (between brackets) the
exposure time in seconds. The derived spectral types are given in the
sixth column. Spectral types marked with a '*' are less secure than
the average, because of poor signal to noise, presence of double lines
or absence of one of the two spectral regions. Stars whose spectral
types are  given as ``($-$)'' were re-observed with the 1.93-m
telescope and their spectral types appear in
Table~\ref{tab:ohp2}. References 
for the photometric values in columns 7 and 8 are
(h) \citet{haug}, (hi) \citet{hil56}, (w) \citet{wram76} and
(t) $Tycho$ photometry \citep{tycho}.
$V$ values marked with a ``$^{*}$''
indicate  formal errors larger than $0.05\:$mag in {\em Tycho} photometry.}
  \begin{center}
\begin{tabular}{lccccccccc}
LS & BD/HD  & $l$ & Position 1 & Position 2& Spectral & $V$ &  $(B-V)$ &
Ref. & $DM$\\
Number& number && & & Type &  & &\\
\hline
&&&&&&&\\
III $+46\degr$19 & $-$ & $85\fdg3$& 18/09 (1800) & 19/09 (1800) & B1V+&11.38 & 0.61&w&12.0\\
III $+45\degr$54 & $-$ & $87\fdg9$ & 21/09 (2475) & 20/09 (2515) & B3Ib &11.49& 0.39&w&15.7\\
III $+46\degr$50 & $-$ &$88\fdg8$ & 21/09 (1800) & 20/09 (1800) & O9.5III &10.69 & 0.72&w&13.1\\
III $+53\degr$32 & $-$ &$99\fdg8$ & 18/09 (1200) & 19/09 (1200) & B2Ib& 10.69& 0.40&w &14.8\\
III $+52\degr$31 & $-$ &$100\fdg0$  & 18/09 (1200) & 19/09 (1200) & ($-$)&10.79& 0.23&w & ($-$) \\
III $+55\degr$19 & $-$ & $100\fdg5$ & 18/09 (1200) & 19/09 (1200) &
B0III$^{*}$+? &10.64& 0.63 &w &13.3\\
III $+59\degr$19 & $-$ & $103\fdg4$& 21/09 (2400) & 20/09 (2400) &B1.5III$^{*}$ &11.02&0.56&w&12.6\\
III $+55\degr$88 & $-$ & $103\fdg8$& 31/07 (1000) & 01/08 (1500) & B0.2IV & $11.15^{*}$&0.28 &t& 14.0\\
III $+57\degr$112 & $-$ &$108\fdg5$ & 18/09 (1500) & $-$ & ($-$) &10.55& 0.81 &w &($-$)\\
III $+58\degr$70 & $-$ &$108\fdg8$ & $-$ & 19/09 (1800)& ($-$)&11.32 &0.45 &w &($-$)\\
I $+60\degr$33 &$-$ &$115\fdg3$ & 18/09 (1800)& 19/09 (1800) & B0III+& 10.97&0.61&h &13.7\\
I $+60\degr$37 &$+59\degr$2779 &$115\fdg4$ & 18/09 (1500)& 19/09 (1500) & B0.2V& 10.48 &0.78 & h& 11.3\\
I $+62\degr$23 & $+62\degr$2296A & $115\fdg8$ & 18/01 (600)& 19/01 (720) & B2.5Ia&8.64& 1.07 & h & 12.3\\
I $+60\degr$46 & $+60\degr$2631 & $115\fdg8$ & 18/01 (600)& 19/01 (750) & B0.5III &9.65 &0.51&h &12.4\\
I $+60\degr$51 & $+60\degr$2635 &$115\fdg9$ &  31/07 (900) & 01/08 (900) & ON7III(f) & 10.13 & 0.46 & h & 13.5\\
I $+63\degr$22 & $+63\degr$2084 &$117\fdg2$ & 18/01 (600) & 19/01 (720) & B2III &9.20 &0.25& t &11.5\\
I $+60\degr$77 &225146 &$117\fdg2$& 18/01 (600)& 19/01 (600) & O9.7Ib &8.59 & 0.38 &h &12.7\\
I $+62\degr$59 & $+62\degr$2353 &$117\fdg4$ & 18/01 (720) & 19/01 (750) & B0.5III &9.88 & 0.24& h &13.5\\
I $+63\degr$57 & $+63\degr$18&$119\fdg0$ & 18/09 (1000)& 19/09 (1000) & B0.2IIIe &10.24 & 0.53 & w &$\ga13.0$\\
I $+61\degr$153 & $-$ &$119\fdg9$ & 18/09 (1800)& 19/09 (1800) & B2III & 11.13 & 0.14 &h &13.8\\
I $+62\degr$107 &$+61\degr$90a&$120\fdg3$ & 18/09 (1000)& 19/09 (1200) & B0.5V+&10.20 & 0.18 &h &12.7\\
I $+62\degr$113 &$+62\degr$89 & $120\fdg5$& 18/09 (1000)& 19/09 (1200) & B1.5Ve&10.15 & 0.49 &h &$\ga 10.8$\\
I $+60\degr$247&$+59\degr$524 &  $135\fdg9$& 18/09 (1800)& 19/09 (1800) & B0.7II-III& 10.95 & 0.63 &t &13.1\\
I $+57\degr$138 & $-$ &$146\fdg3$ & 20/01 (1000) & 19/01 (1000)& O7V &10.09&0.28&h&13.2\\
I $+57\degr$139 &$+56\degr$864 &$146\fdg3$& 20/01 (900) & 19/01 (900) & O6V+ &9.67&0.28&h&$>13.0$ \\
I $+56\degr$97 & BD $+56\degr$866 &$146\fdg4$& 20/01 (1200) & 21/01 (1200)&O9V&10.33 &0.34 & h &12.8\\ 
I $+55\degr$55 & $+55\degr$837 &$147\fdg0$& $-$ & 25/02 (1500) & B1Ib$^{*}$&9.59&0.69&h&12.9\\
I $+56\degr$99 & 237211 &$147\fdg1$ & 18/01 (720) & 19/01 (700)& O9.5Iab& 9.00 &0.49&h&13.0\\
V $+56\degr$56 & 25914 & $147\fdg4$ & 20/01 (750) & 22/05 (1000)& B5Ia&7.99&0.60& hi &12.9\\
V $+55\degr$11 & 237213 &$147\fdg6$& 20/01 (900) & 22/05 (900)& B6Ia &8.72 &0.77&hi &13.2\\
V $+56\degr$59 & $-$ &$149\fdg7$& 20/01 (1800) & $-$ & B1V$^{*}$+&10.92  &0.25&t&12.6\\
V $+56\degr$60 & $-$ &$149\fdg8$ &20/01 (1800) & 21/01 (1550)& B2.5III &$11.51^{*}$&0.24&t&13.7\\ 
V $+53\degr$22 & 232947 &$151\fdg6$& 18/01 (720) & 19/01 (750)& B0Ia &9.32 &0.64&hi&13.3\\
V $+40\degr$47& BD $+39\degr$1328 &$169\fdg1$& 18/01 (750) & 19/01 (1000) &O9Ibf&9.85&0.57&hi&13.6\\
V $+33\degr$36 & HD 243827 &$174\fdg3$&18/01 (900) & 19/01 (1200)& B0.2III &10.61&0.45&hi&13.6 \\ 
V $+25\degr$20 & $-$ &$185\fdg5$& 18/01 (1800) & 19/01 (1200)& B0.3V &$10.40$ &0.21&t&13.0\\
V $+22\degr$5& 248893 &$186\fdg8$& 20/01 (1000) & 25/02 (1200)& B0III &9.75&0.25&t&13.4\\
V $+22\degr$38 & $-$ &$188\fdg8$& 20/01 (2000) & 25/02 (2400)& B0.2III$^{*}$(+) &10.76&1.01&h&12.0\\
V $+20\degr$20 & 252535 &$190\fdg0$& 18/01 (600) & 19/01 (1000)& B1.5V(+) &10.09 &0.19 &h&11.6\\
\end{tabular}
\end{center}
 \label{tab:ohp1}
\end{table*}

Already at the time when the first basic picture appeared, it was
noted that several young star clusters 
in the Anticentre direction were located at heliocentric distances of
5-6 kpc \citep{mav75}, far beyond the expected position of the Perseus
Arm, and it was suspected that they were tracers of a more external
($+$II) arm. More recent work has confirmed the long distances to 
clusters such as Bochum 2 \citep{mac95}  or Dolidze 25
\citep{len90} -- both at ($l\approx212\degr$). In addition,
\citet{tm93} found 6 other young star groupings associated to
\ion{H}{ii} regions in this area, lying at distances in the $5-9$ kpc
range. 
Recent distance determinations indicate that the much larger open
cluster \object{NGC 1893} ($l=173.6\degr$), previously believed to be
the core of the Aur OB2 association, is also located 
at a heliocentric distance of $\sim 6$kpc \citep{chopi}, suggesting
that this star-forming region also belongs to the Outer Arm.

Molecular clouds definitely delineating an Outer Arm are readily visible
in CO surveys all over the first Galactic quadrant, where they display
radial velocities clearly distinct from those of Perseus Arm clouds
\citep{dam01}. This Outer Arm, the first spiral arm encountered
beyond the Perseus Arm, is generally referred to as the Cygnus Arm. In
most models,  it is assumed to be the 
continuation of the Norma Arm and therefore a main spiral
feature. However, its definition is much less clear in the second
quadrant, where the Cygnus Arm must lie basically beyond the Perseus
Arm and its tracers are sparse.  

The presence of elements very clearly belonging to the Cygnus Arm in
the first Galactic quadrant and again in the Anticentre region and
third quadrant is naturally suggesting that the Cygnus Arm must also
exist in the second quadrant. Evidence for \ion{H}{i} structure 
beyond the Perseus Arm has been collected from radio observations
(e.g., \citealt{kul82}; see also references in \citealt{ww91}).
\citet{rus03} finds several star-forming regions likely
associated with the Cygnus Arm, but she is unable to decide whether
these are actually organised into an arm structure. It has been
suggested \citep[e.g.,][]{qui02} that the morphology of the outer
Milky Way could be essentially flocculent. 

\citet{kw89}, collecting a large amount of data from the literature, 
carried out a statistical analysis of the distribution with distance of
spiral tracers in the second quadrant. They concluded that the
Cygnus Arm is well defined in the sense that, when considering spiral
tracer counts with distance, a statistically
significant gap exists between the Perseus Arm and a second maximum at
$\sim 5$ kpc. Though the results of \citet{kw89} are highly significant,
their procedure is affected by a number of factors that limit their
validity: they use data from a large variety of inhomogeneous sources,
estimate most of their distances from photometric data only and do not
have sufficient distant tracers to attempt the study of their spatial
distribution (for example, at $l=90\degr$, the Cygnus Arm is expected
to be at $6-7$ kpc, while at $l=150\degr$, it should be at $4-5$
kpc; cf. \citealt{val02}).
 
In an attempt to obtain a more uniform data sample, we have started a
programme to study the distance to different spiral tracers. In this
first paper, spectroscopic parallaxes are derived for a sample
of OB stars spanning the whole of the Second Galactic Quadrant,  whose
photometric indices are reported in the literature 
to indicate large distances. More detailed spectroscopic and
photometric studies of individual young open clusters and OB
associations will be presented in successive papers.

\section{Observations}

Most observations have been carried out with the {\em Aur\'{e}lie}
spectrograph on the 1.52-m telescope at the Observatoire de Haute
Provence (OHP) during three dedicated runs on 18th-24th September
2001, 18th-22nd January 2002 and 25th-28th February 2002. In addition, 
one single object (\object{LS I +60\degr51}) was observed during a
different run on July 31st/August 1st 2001. The
spectrograph has been equipped with grating \#3 (600 ln mm$^{-1}$) and
the Horizon 2000 $2048\times1024$ EEV CCD camera (see \citealt{gil94}
for a description of the instrument). In the classification region,
this configuration gives a dispersion of 0.22\AA/pixel (resolving
power of approximately 7000), covering a wavelength range of $\approx
440$\AA. It is therefore necessary to observe two wavelength regions
in order to span the classical classification region.

\begin{table*}[ht]
\caption{Distant OB stars observed from the 1.93-m OHP
telescope. \object{LS I $+60\degr$78} is also known as \object{BD
$+60\degr$2664} and \object{LS II $+28\degr$12} is listed as 
\object{HD 332907}. References
for the photometric values in columns 6 and 7 are (tu) \citet{tur80}
(h) \citet{haug}, (w) \citet{wram76}, (w2) \citet{wram81} and
(t) $Tycho$ photometry \citep{tycho}.}
  \begin{center}
\begin{tabular}{lcccccccc}
LS & $l$ & Date& Exposure &Spectral & $V$ &  $(B-V)$ & Ref. & $DM$\\
Number& & &Time (s) &  Type &  & &\\
\hline
II $+28\degr$12 & $64\fdg4$ & 06/07 & 650 &B0.7II& 10.91&0.45&tu&14.2\\
III $+47\degr$39 & $91\fdg3$ & 07/07 &1200&B1II& 11.76 &0.90& w &13.5 \\
III $+47\degr$41 & $91\fdg3$ & 07/07 &1200&B1III& 12.44 &0.71& w &13.9\\
III $+52\degr$17 & $98\fdg5$ & 06/07 &1200&B2IV& 12.25 &0.43 & w & 13.4\\
III $+52\degr$19 & $98\fdg6$ & 06/07 &1200&BN1IIIe& 12.04 &0.44 & w&$\ga14.3$\\
III $+52\degr$31 & $100\fdg0$ & 07/07 &750& B1.5Ve &10.79& 0.23 &w&$\ga12.2$\\
III $+52\degr$32 & $100\fdg0$ & 07/07 &750&B2IV&10.43  &0.06& w &12.7\\
III $+55\degr$20 & $100\fdg4$ & 07/07 &900&BN0IV& 11.07  & 0.77 &w &12.6\\
III $+59\degr$29 & $107\fdg3$ & 06/07 &750&B0.7III& 11.17 & 0.66&w &13.2\\
III $+57\degr$111 & $108\fdg4$ & 06/07 &900& B0.2III& 10.32 & 0.61  & t &12.8\\
III $+57\degr$112 & $108\fdg5$ & 06/07 &900&B0.3III& 10.55 & 0.81&w &12.5\\
III $+58\degr$70 & $108\fdg8$ & 06/07 &1200&B0V& 11.32 &0.45 & w&13.3\\
III $+58\degr$71 & $108\fdg9$ & 06/07 &1200&B0.3V& 12.12 &0.43 & w2&14.0\\
I $+59\degr$10 & $113\fdg9$ & 07/07 &900&B0.2III& 10.81 &0.80& h &12.7\\
I $+62\degr$43 & $116\fdg3$ & 07/07 &1800&B2IV& 12.34 &0.37& h &13.6\\
I $+62\degr$44 & $116\fdg3$ & 07/07 &1800&B2.5IV& 12.56 &0.47& h &13.6\\
I $+60\degr$78A & $117\fdg2$ & 06/07 &750&B1V& 10.62 &0.37 & h&12.0\\
I $+60\degr$78B& $117\fdg2$ & 06/07 &750&B1.5V& 10.65 &0.33 & h&11.7\\
\end{tabular}
\end{center}
 \label{tab:ohp2}
\end{table*}

For this programme, the two regions selected were centred at
$\lambda=4175$\AA\ and $\lambda=4680$\AA. In principle, all objects
were intended to be observed in both 
regions, resulting in coverage in the $\lambda\lambda 3950 - 4900$\AA\
range, with a small gap ($\sim 60$\AA) around $\lambda 4425$\AA. No
strong photospheric lines are found in the gap, but the strong diffuse
interstellar line (DIB) at $\lambda 4428$\AA, which is a good
indicator of the reddening, is lost.
However, due to the weather, only the equivalent of 6 whole nights
could be used for observing and a small number of objects were only
observed in one position. Fortunately, for most spectral types in the
range considered any of the two regions allows an accurate spectral
classification. 

 The complete log of observations at the 1.52-m OHP is given
in Table~\ref{tab:ohp1}. In addition to the stars listed, Position 2
observations were obtained on 20th Sep 2001 for \object{LS I
+62\degr99} and \object{LS I +62\degr119}. However, since the
Signal-to-Noise Ratio (SNR) of
these spectra is very low (due to thick veiling) and both objects
appear to be Be stars (as H$\beta$ seems in emission), no spectral
types have been derived.

Several stars were observed on the nights of 6th and 7th July 2002
using the 1.93-m telescope at the OHP, France. The telescope was
equipped with the long-slit  
spectrograph {\em Carelec} and the EEV CCD. The 1200 ln\,mm$^{-1}$
grating was used, giving a nominal dispersion of $\approx
0.45$\AA/pixel over the range 4000\,--\,4900\AA. The spectral
resolution is  $\approx 1.5$\AA. The observations from the 1.93-m OHP
are listed in Table~\ref{tab:ohp2}.

Finally spectra of a few objects in the Galactic Anticentre were
obtained on the nights of 
December 5-7th 2001 using the 
Andalucia Faint Object Spectrograph and Camera (ALFOSC) on the 2.6-m
Nordic Optical Telescope (NOT), in La Palma, Spain. The telescope was
equipped with a thinned $2048\times2048$ pixel Loral/Lesser
CCD. Spectra were taken with grism \#7 and the slit width was set to
$1\farcm0$, resulting in a spectral resolution of $\approx 6.6$\AA.
The log of these observations is given in Table~\ref{tab:not}.

\begin{table*}[ht]
\caption{Distant OB stars observed from the NOT. References are (h)
\citet{haug}, (m) \citet{mc67} and
(t) $Tycho$ photometry \citep{tycho}. Note that M $+22\degr$178 is
taken from \citet{mc67} and does not have an LS number. Photometry
from \citet{mc67} is likely to be rather inaccurate. Spectral types
for these objects are rather less accurate than for stars in
Tables~\ref{tab:ohp1} and~\ref{tab:ohp2}.}
  \begin{center}
\begin{tabular}{lcccccccc}
LS & $l$ & Date& Exposure &Spectral & $V$ &  $(B-V)$ & Reference & $DM$\\
Number& & &Time (s) &  Type &  & &\\
\hline
V $+22\degr$6 & $186\fdg3$ & 07/12 &600&B1V&10.96&0.34 &t&12.4\\
V $+23\degr$8 & $186\fdg4$ & 05/12 &700&O8Iaf&10.83&1.25&m&12.5\\
M $+22\degr$178 &$186\fdg6$  & 05/12 &750& B2.5III$^{*}$&12.33 &1.04&m&12.1\\
V $+22\degr$50 & $189\fdg7$ &07/12 &600&B2IVe&11.47&0.22&h&13.2\\
V $+19\degr$2 &$191\fdg5$ & 07/12 &900& B1.5III$^{*}$&12.06& 0.58&h&13.5\\
\end{tabular}
\end{center}
 \label{tab:not}
\end{table*}

All the spectroscopic data have been reduced with the {\em Starlink}
packages {\sc ccdpack} \citep{draper} and {\sc figaro}
\citep{shortridge} and analysed using {\sc figaro} and {\sc dipso}
\citep{howarth}. 

\section{Target selection and methodology}
\label{sec:method}

For this programme, we have selected relatively bright stars given in
the literature as candidates to distant OB stars because of their
colours. These objects concentrate in a few interesting regions of the
second Galactic quadrant. The location of Cygnus Arm stars in the first
quadrant is to be a very difficult task, because of the large distance
and strong absorption due to the Orion Spur.
The stars observed have been mainly selected
from \citet{haug} and \citet{wram76}, but other sources have been used
(mostly those given in \citealt{kw89}). The exposure times needed with the
configuration used at the OHP 1.52-m telescope makes advisable not to
observe stars with apparent magnitudes
$B\ga 12.5$, though a few slightly fainter objects have been observed
with the larger telescopes. 

In any case, the limit imposed by the instrumentation is approximately
coincident with the magnitude limit of the Luminous Star (LS)
catalogues (though they are not complete to this depth),
beyond which there are very few available sources for
candidates -- fainter OB stars have not in general been identified as
such and photometric data are not available. Therefore the sample
observed cannot at all intend to represent in any significant way
the population of massive stars in 
the Cygnus Arm, but rather consists only of a few glimpses through
windows of low absorption in the Perseus Arm. 

In order to derive spectroscopic parallaxes, accurate photometry (at
least in $B$ and $V$) is necessary. From the derived spectral type and
measured $(B-V)$, one can then calculate $E(B-V)$, making use of
tabulated calibrations of the intrinsic colour $(B-V)_{0}$
corresponding to each spectral type. Assuming then a standard value
for the ratio of total to selective absorption $R$, a value for the
visual extinction $A_{V}$ is derived. The resulting magnitude
corrected for reddening can then be used to derive the distance
modulus ($DM$), making use of a calibration of absolute magnitude for
a given spectral type.

There are many sources of uncertainty in this procedure. In principle,
if the observational data are accurate, the two main sources of
uncertainty are the value of $R$ and the intrinsic dispersion in the absolute
magnitude calibration. For a given spectral type, a dispersion of
several tenths of a magnitude in the absolute
magnitude calibration is normal among O-type and
early B-type stars \citep[cf.][and references therein]{wal02}. The
value of $R$ can be 
locally very different from the Galactic average $R=3.1$
\citep{fit99}. Both 
effects lead to a rather large dispersion in the derived $DM$, but
assuming that deviations from standard values are random in nature,
they can be compensated by observing a high number of stars. 

Systematic effects are more difficult to correct. The absolute
magnitude calibration for O-type stars is still
debated. From an analysis using plane-parallel, non-LTE, pure H and He
hydrostatic models, \citet{vac96} derived a higher temperature scale
and brighter magnitudes than previous calibrations (e.g.,
\citealt{hme84}). More recent models, including line blanketing,
sphericity and mass loss (e.g., \citealt{mar02,her00}) support lower
effective temperatures and similar magnitudes to the calibration of
\citet{vac96}. On the other hand, \citet{bg02}, from an analysis of
{\em FUSE} spectra, find even lower effective temperatures and lower
magnitudes. 

As these difficulties have still to be solved, in this
work, we will simply have to use a consistent magnitude scale. We will
use the calibration of  \citet{hme84} for spectral types B2 and later
and the calibration of \citet{vac96} for spectral types O9 and
earlier. As there is an important discrepancy between these two
calibrations around spectral type B0, where the absolute magnitude
changes rather steeply with spectral type, we have opted for an interpolated
calibration, which is given in Table~\ref{tab:cal}. This calibration
is similar, though not identical, to that used by \citet{rus03}. We
assume slightly brighter magnitudes for stars of spectral types
O9.5-B0.5, after \citet{vac96}.

The calibration
for intrinsic colours used is that of \citet{weg94}. We have
interpolated over intermediate spectral types not tabulated in this
work, and assumed that luminosity class IV stars have the same
intrinsic colours as main sequence stars.

In order to prevent systematic observational effects, a
homogeneous dataset and analysis is fundamental for accurate
distances. In this work, the spectroscopic dataset is homogeneous and
the spectral classification has been done in a consistent way,
reducing the possibility of artificial dispersion. A homogeneous
photometric dataset, however, is not available in the literature. 
This turns out to be the main source of systematic uncertainty in our
results.

\begin{figure}
\begin{picture}(250,260)
\put(0,0){\includegraphics{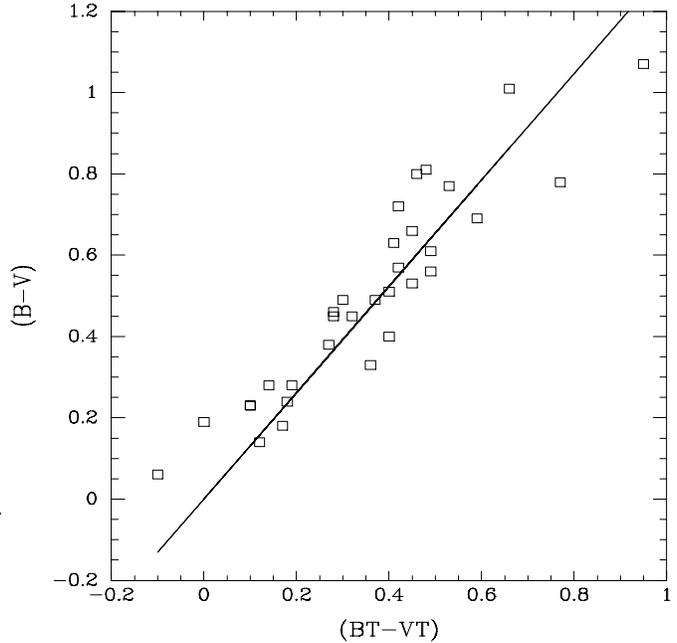}}
\end{picture}
  \caption{Lineal regression between photoelectric $(B-V)$ and {\em
   Tycho} $(B_{T}-V_{T})$ for 30 stars in our sample for which both
   kinds of data exist. The lineal fit provides a relationship
   $(B-V)=1.31(B_{T}-V_{T})$, with a correlation coefficient
   $R=0.89$. This is very different from the standard relation for
   {\em Tycho} photometry, derived for unreddened stars.}
   \label{fig:tychojo}
\end{figure}

For a large fraction of our sample, photometry is available
from the {\em Tycho} catalogue \citep{tycho}. It was therefore our
initial intention to use this dataset as primary source for
photometry. 
However, comparison of all existing photometric data for several
objects showed very good agreement between different photoelectric
measurements and large and {\em systematic} discrepancies between
photoelectric and {\em Tycho} photometry. 

\begin{table}[t]
\caption{Absolute magnitude calibration used here. For stars earlier
than B0, it is based on the calibration of \citet{vac96}. For B
spectral types, the calibration of \citet{hme84} has been used. In the
B0-B1 interval, where the calibrations are very discordant, an
interpolation has been made. For the only star with luminosity class
Iab in the sample, \object{HD 237211} (O9.5Iab), $M_{V}=-6.3$ has been
adopted.}
  \begin{center}
\begin{tabular}{lcccccc}
& V & IV & III & II & Ib & Ia\\
\hline
O6 &$-5.1$&$-$ &$-5.8$&$-$& $-$& $-6.4$\\
O7 &$-4.9$&$-$ &$-5.7$&$-$& $-$& $-6.5$\\
O8 &$-4.7$&$-$ &$-5.6$&$-$& $-$& $-6.5$\\
O9 &$-4.4$&$-$ &$-5.5$&$-$& $-6.3$& $-6.5$\\
O9.5/7 &$-4.3$&$-$ & $-5.4$& $-$&$-6.0$&$-6.6$\\
B0 &$-4.2$&$-4.7$&$-5.3$&$-5.8$&$-6.0$&$-6.6$\\
B0.2/3 &$-4.0$&$-4.6$&$-5.1$&$-5.6$&$-6.0$&$-6.7$\\
B0.5 &$-3.8$&$-4.5$&$-5.0$&$-5.5$&$-6.0$&$-6.9$\\
B0.7 &$-3.5$&$-4.2$&$-4.8$&$-5.3$&$-6.0$&$-7.0$\\
B1 &$-3.2$&$-3.8$&$-4.3$&$-5.1$&$-6.0$&$-7.0$\\
B1.5 &$-2.8$&$-3.3$&$-3.9$&$-5.0$&$-5.8$&$-7.2$\\
B2 &$-2.5$&$-3.1$&$-3.7$&$-4.8$&$-5.8$&$-7.4$\\
B2.5 &$-2.0$&$-3.1$ &$-3.5$&$-4.8$&$-5.8$&$-7.4$\\
B3 &$-1.6$&$-2.5$&$-3.0$&$-4.7$&$-5.8$&$-7.2$\\
\end{tabular}
\end{center}
 \label{tab:cal}
\end{table}

It was observed that the $(B-V)$ colour was systematically larger in
photoelectric photometry than when the standard transformation
\begin{equation}
(B-V)=0.85(B_{T}-V_{T})
\label{eq:tycho}
\end{equation}
from {\em Tycho} photometry \citep{tycho} is applied. This effect is
illustrated in Fig.~\ref{fig:tychojo}. When a lineal fit is applied to
the 30 objects in our sample for which both {\em Tycho} and
photoelectric photometry exists, a transformation
\begin{equation}
(B-V)=1.31(B_{T}-V_{T})
\label{eq:ours}
\end{equation}
is derived, clearly very different from Eq.~\ref{eq:tycho}.

The existence of this difference is not surprising as
Eq.~\ref{eq:tycho} has been derived from a sample that mostly consists
of nearby unreddened stars, while our sample wholly consists of
reddened intrinsically blue stars. As a matter of fact, it appears
from our small sample that the difference between photoelectric
$(B-V)$ and the values derived from Eq.~\ref{eq:tycho} increases with
increasing $E(B-V)$ (calculated assuming the photoelectric
values). This point should be investigated with a larger sample of
stars for which reliable spectral types exist.

In view of this discrepancy, we have opted for taking \citet{haug} as
our prime source of 
photometry. Objects not observed by \citet{haug} have been taken from
\citet{wram76}, as this author has transformed his measurements to be
in the same system used by \citet{haug} and \citet{hil56},
cf. \citet{sw77}. For a few stars, we have taken 
photoelectric photometry from other sources, listed in
Tables~\ref{tab:ohp1} and~\ref{tab:ohp2}. Only when we have been
unable to find {\em any} photoelectric photometry in the literature,
have we resorted to {\em Tycho} photometry. For two stars in
Table~\ref{tab:not}, no photoelectric or modern photometry existed. We
have had to resort to photographic photometry from
\citet{mc67}. Comparison of values from \citet{mc67} for other stars with
more modern photoelectric photometry shows that these photographic values
may be in error by several tenths of a magnitude.

\section{Results}

For all stars observed accurate spectral types have been derived. For
the majority of targets, the signal-to-noise ratio is sufficiently
high to allow, at this resolution, an accuracy better than the
subtype. Classification has been carried out by direct comparison to
MK standards observed at similar (or slightly lower) resolution and
the digital atlas of \citet{waf}. We followed the guidelines for
classification from \citet{waf}, complemented by the spectral atlas of
\citet{len92} for the supergiants. For O-type stars, the procedure of
\citet{waf} has been complemented by the quantitative relations of
\citet{mat}, based on Conti's scheme. In all cases, both methods have
given excellent agreement, demonstrating once again their consistency.

Once a spectral type was obtained, we calculated a distance modulus
following the procedure described in the previous section. The
distribution of stars with distance has been studied by dividing the
sample into several individual regions. A histogram plot showing the
distribution in three of those areas is shown in Fig.~\ref{fig:histo}.

\subsection{Vulpecula}

In the Vulpecula and Cygnus region, the study of the Cygnus Arm is
complicated by the fact that distances are very large and the Orion
Spur runs almost parallel to the line of sight, resulting in
heavy obscuration and poorly defined distances. In his study of
\object{Vul OB2}, \citet{tur80} determined a dereddened distance
modulus $DM=13.2$, placing this association at 4.4 kpc, in good
agreement with the expected position of the Perseus Arm
\citep{val02}. We searched for candidates for Cygnus Arm stars in the
works of \citet{tur80} and \citet{for85}, but few appeared convincing.

\begin{figure*}
\begin{picture}(500,280)
\put(0,0){\includegraphics{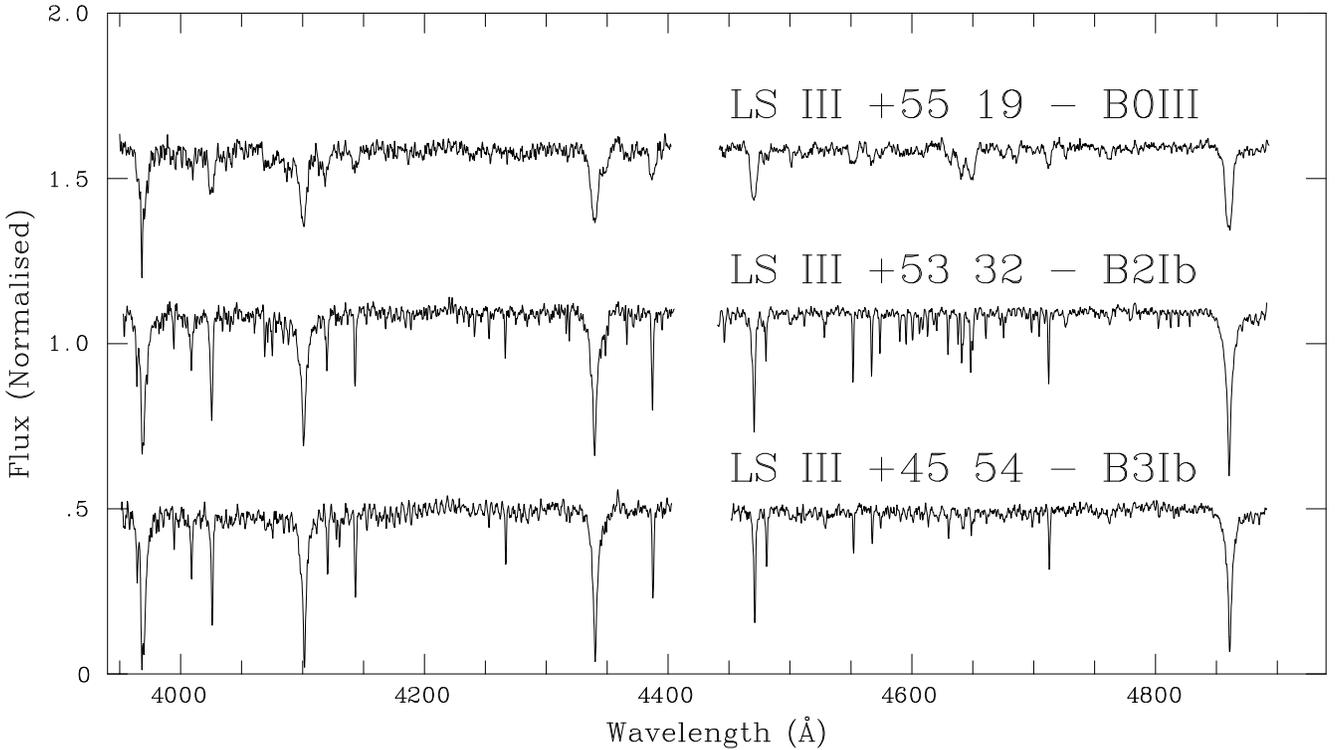}}
\end{picture}
  \caption{Likely tracers of the Cygnus  Arm in the $l = 85\degr - 105
   \degr$ range
   observed with the OHP 1.52-m telescope. The small gap around
   $\lambda = 4420$\AA\ indicates the division between the two
   poses. \object{LS III +53\degr32} and \object{LS III +45\degr54}
   have spectroscopic distances well in excess of $d=7\:$kpc.}
   \label{fig:first}
\end{figure*}

The only star we have observed is \object{LS II $+28\degr$12}. The
derived spectral type B0.7II implies $DM=14.2$ using the photoelectric
photometry from \citet{tur80}. This value is rather larger than the
distance to the Perseus Arm, but still shorter than the $DM\approx14.8$
compatible with the $\approx 9$ kpc distance to the Cygnus Arm
in this direction \citep{val02}. The case therefore remains
inconclusive.

\subsection{The Cepheus region}

The \object{Cep OB1} association extends, according to
\citet{humphreys}, from $l= 98\degr$ to $l=108\degr$. \citet{gs92}
list numerous members, but warn that it could well really consist of
two independent associations, centred on the Eastern and Western sides
of this range \citep[cf.][]{mof71}. The reality of this association
has also been disputed by other authors
\citep[e.g.,][]{mae95}. Assuming it is a single association,
\citet{gs92} derive a $DM=12.2$.

We have observed 19 stars over the longitude range
$85\degr<l<110\degr$, 14 of which lie within the traditional limits of
\object{Cep OB1}.
The spectrum of \object{LS III +55\degr20} seems to be moderately N 
enhanced and rather C deficient. \object{LS III +52\degr19}, apart
from being a Be star, also seems to be C deficient. Another Be star in
this range is \object{LS III +52\degr31}. The distances to Be stars
should be taken as lower limits,
since part of the $E(B-V)$ excess will be due to circumstellar emission
and not interstellar absorption and also because Be stars tend to be
slightly brighter than normal stars of similar spectral types. These two
factors are unlikely to result in a difference much larger than $0.3$
mag. 

\begin{figure*}
\begin{picture}(500,300)
\put(0,0){\includegraphics{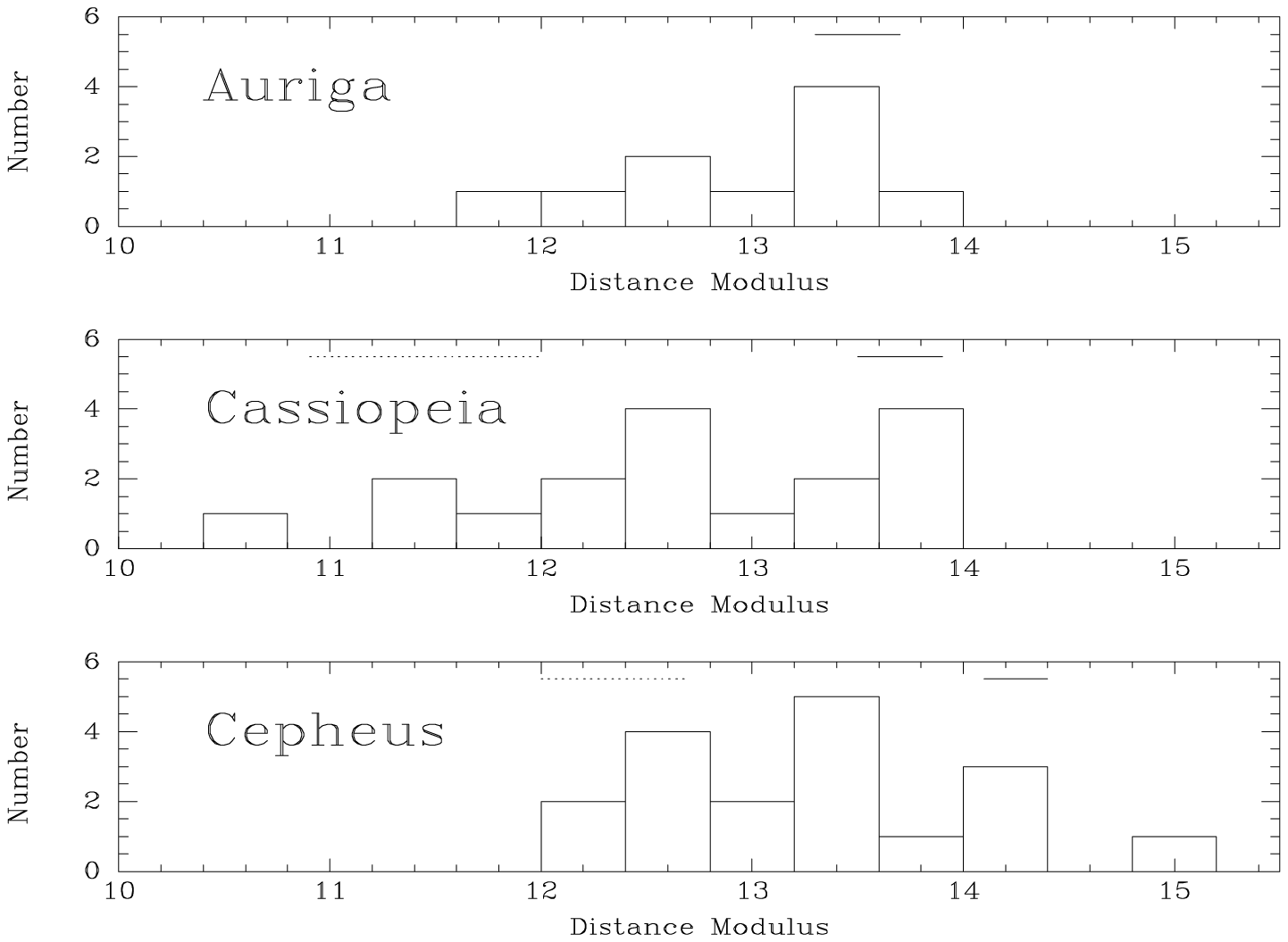}}
\end{picture}
  \caption{Histogram showing the spread of stars with distance modulus
   in three of the regions considered in the text. Stars are arbitrarily
   grouped in 0.4-mag bins. Dotted lines on top of the graphs
   show the traditionally-accepted extent of the Perseus Arm in the
   Cepheus and Cassiopeia regions, while continuous lines show the
   estimated position of the Cygnus Arm, from the model of
   \citet{val02}. In all cases, the arms are assumed to have a width
   of 1 kpc. In all three regions there are obvious
   concentrations of stars at distances compatible with the Cygnus
   Arm. Note that the distance to \object{LS III $+45\degr$54} is too
   large to fit in the figure. Also note that all stars in our sample
   were selected because their photometric indices suggested that they
   were very distant objects. As the original intention was not to include
   any Perseus Arm objects, the distribution shown here is
   not representative {\em at all} of the total distribution of stars.}
   \label{fig:histo}
\end{figure*}

Seven stars turn out to have $DM\leq12.8$. The distances to \object{LS III
+46\degr19} and \object{LS III +52\degr31} must be considered lower
limits, as they are respectively a double-lined spectroscopic binary
and a Be star. The distances to the other five objects cluster quite
tightly, with an average $\overline{DM}=12.64\pm0.10$ (errors in
averages indicate 1-$\sigma$ deviations). The true
distance moduli to  \object{LS III +46\degr19} and \object{LS III
+52\degr31} are likely to be compatible with this value. 
The corresponding distance $d=3.4$ kpc is moderately larger than the
distance derived to \object{Cep OB1} by \citet{gs92}, but clearly
consistent with the distance to \object{NGC 7380}, believed to
be its nuclear cluster, derived by other
authors. For example, photometric determinations give $DM=12.5$
\citep{baa83,cha94}, while \citet{mas95} find $DM=12.9\pm0.1$ from 
spectroscopic parallaxes of 10 stars. Therefore this seven stars in
our sample appear to be members of \object{Cep OB1} and hence be
located on the Perseus.

Six stars have $13.1\leq DM \leq 13.5$. The average is
$\overline{DM}=13.3\pm0.1$. There appears to be a clear gap between
this group and the Perseus Arm group. Moreover, this is the most
obvious concentration of stars in this region (see
Fig.~\ref{fig:histo}), though the corresponding distance $d=4.6$ kpc
does not fit well with the position of any Galactic arm in current
models, unless the Perseus Arm is considered to be very broad.

The remaining six stars have $DM\geq13.9$. Again there is a significant
gap with respect to the previous group (see
Fig.~\ref{fig:histo}). Three stars cluster 
very tightly around $DM=14.0$, but adding \object{LS III +52\degr19}
and \object{LS III +53\degr32}, we obtain an average
$\overline{DM}=14.2\pm0.3$. The corresponding distance $d=7$ kpc is
consistent with the expected position of the Cygnus Arm.
 
The $DM=15.7$ to \object{LS III $+45\degr$54} is remarkably large,
implying $d=13.8\:$kpc. The
photometry is not in error, as \citet{dri75}, gives $V=11.52$,
$(B-V)=0.38$, in very good accordance with Wramdemark's
values. Reducing the luminosity class to B3II would lower the $DM$ to
14.6, more in line with other tracers of the Cygnus Arm, but the
spectrum looks that of a supergiant (see Fig.~\ref{fig:first}).

\begin{figure*}
\begin{picture}(500,280)
\put(0,0){\includegraphics{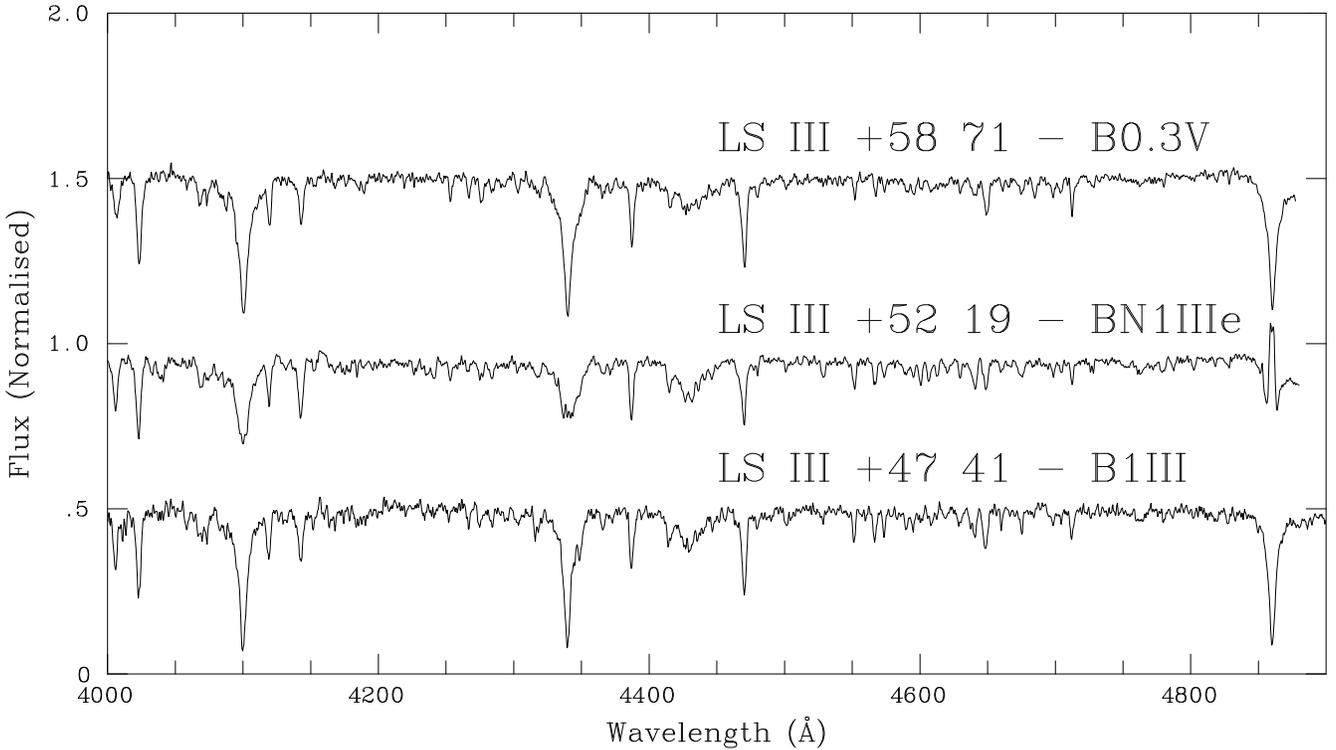}}
\end{picture}
  \caption{Likely tracers of the Cygnus Arm in the region of
   \object{Cep OB1} observed  with the OHP 1.93-m telescope.}
   \label{fig:ohp}
\end{figure*}

\subsection{Cassiopeia}
\label{sec:cas}

Based on photometry of stars in the LS catalogue, \citet{haug}
suggested that over a rather large region of the Galactic plane,
centred on the association \object{Cas OB5}, there was a
significant population of OB stars at a rather larger distance than
the objects considered to be members. This suggestion was supported by
further photometric \citep{wram76} and spectroscopic \citep{mar72}
studies. 

\begin{figure*}
\begin{picture}(500,280)
\put(0,0){\includegraphics{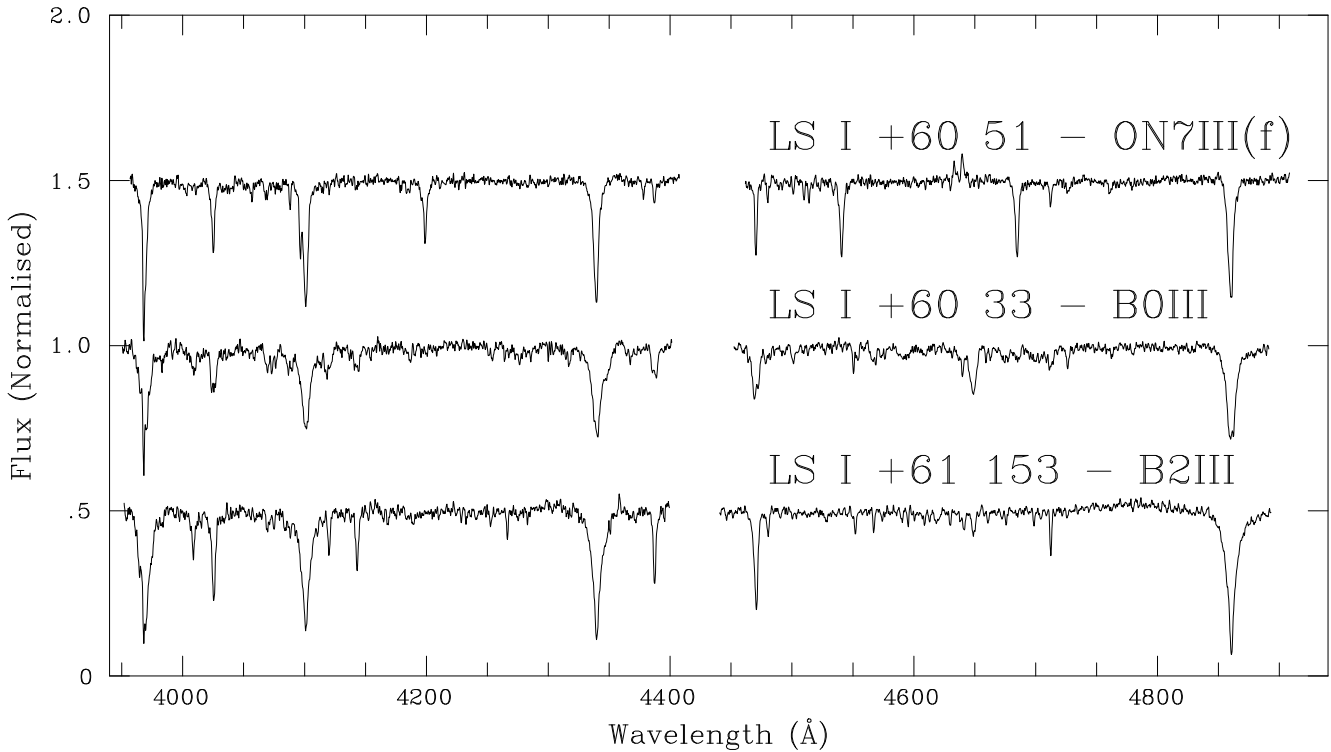}}
\end{picture}
  \caption{Likely tracers of the Cygnus Arm in the region of
   \object{Cas OB5} observed with the OHP 1.52-m telescope. \object{LS
   I +60\degr51} has one of the earliest spectral types observed in
   this survey. The small gap around
   $\lambda = 4420$\AA\ indicates the division between the two
   poses.}
   \label{fig:cas}
\end{figure*}

\object{Cas OB5} is a well populated OB association extending from
$l\approx115\degr$ to $118\degr$ \citep{humphreys}. There is
no obvious core cluster associated, rendering a distance determination
unsure. According to \citet{gs92}, members define a fairly clear
main sequence at $DM=11.5$. Almost coincident with it is \object{Cas
OB4}. It extends from  $l\approx119\degr$ to $122\degr$
\citep{humphreys} and, though sparse, it fits well a $DM=12.2$
\citep{gs92}. Members are of rather late spectral type (earliest
spectral type is O9V) and there is no nuclear cluster.

Over the $112\degr\leq l \leq 121\degr$ range, we have observed 17
stars.  Among the stars observed, \object{LS I $+60\degr$33} and
\object{LS I $+62\degr$107} present evidence   
for a second spectrum of similar spectral type and therefore could
have larger $DM$. \object{LS I $+62\degr$23} = \object{BD
$+62\degr$2296A} has been found to be part of an apparent triple
system, including a 
B0III and a WN4 companion. These companions are unlikely to
contribute much to the total luminosity. The system will be discussed
in a separate work. Two of the stars observed in this region,
namely \object{LS I $+63\degr$57} and \object{LS I $+62\degr$113}, are
Be stars. Their distances are therefore likely to be slightly larger
than calculated.

We find a spread of objects over the $DM = 11-13$ range. A few of them
have distance moduli compatible with the accepted distance to
\object{Cas OB5} (\object{LS I $+60\degr$37}, \object{LS I 
$+63\degr$22} \& \object{LS I $+62\degr$113}), but, if there is any
concentration of stars, it lies at $DM\approx12.6$. Again this
distance does not correspond well to any arm, unless we assume that
the Perseus Arm is very broad. Moreover, most of
the objects in the $DM = 12-13$ range 
(marginally compatible with the distance to \object{Cas OB4})
are located within the longitude range assigned to
\object{Cas OB5}. Therefore we do not find a clear definition between
\object{Cas OB5} and \object{Cas OB4}. 

There is, however, a clear gap with respect to the objects with
$DM\geq13$ (see Fig.~\ref{fig:histo}). Leaving aside the Be star
\object{LS I $+63\degr$57}, we 
find six objects with distance moduli clustering tightly around
$\overline{DM}=13.6\pm0.1$, corresponding to $d=5.3$ kpc. This 
group almost certainly corresponds to the tracers
of the Cygnus Arm over this segment. Of 
particular interest is \object{LS I +60\degr51} (\object{BD 
+60\degr2635}). It is an O-type star of moderate luminosity with
strong \ion{N}{iii} emission (see Fig.~\ref{fig:cas}), for which we
derive a spectral type ON7III(f).

\citet{wal02} has shown that most of the O-type
stars traditionally associated to \object{Cas OB5} have $DM$s
indicating rather larger distances. The revised $DM$s calculated by
Walborn fall in the $12-13$ range, in good agreement with our
sample. Obviously this region merits further study and a dedicated
work is forthcoming. 

\subsection{\object{Cam OB3}}

Beyond $l\approx140\degr$, the Perseus Arm is not very well
defined. This is surprising considering that the spectacular
\object{Per OB1} and \object{Cas OB6} associations extend until this
Galactic longitude, but it seems to be due to a real lack of spiral
tracers and not to local extinction. Orion Spur associations are,
however, present in the $l=140\degr-180\degr$ range, notably
\object{Cam OB1} and \object{Aur OB1}. 

\object{Cam OB3} is a rather diffuse OB association, containing no
known open clusters. Its existence has sometimes be considered
doubtful and it is not included in the work of \citet{gs92}. However,
\citet{haug}, based on $UBV$ photometry of a large number of LS stars,
had considered its existence certain. Using data in the literature for
6 likely members, \citet{humphreys} centred it at ($l=147\degr$,
$b=+3.0$) and derived $DM=12.6$. This is one magnitude
larger than the $DM$'s to \object{Per OB1} and \object{Cas OB6}, the
tracers of the Perseus Arm closest in the sky. Moreover, considering
that the Perseus Arm is running towards its minimum distance to the
Sun in this region, \object{Cam OB3} is clearly too far away to be on
the Perseus Arm.

\begin{figure*}
\begin{picture}(500,280)
\put(0,0){\includegraphics{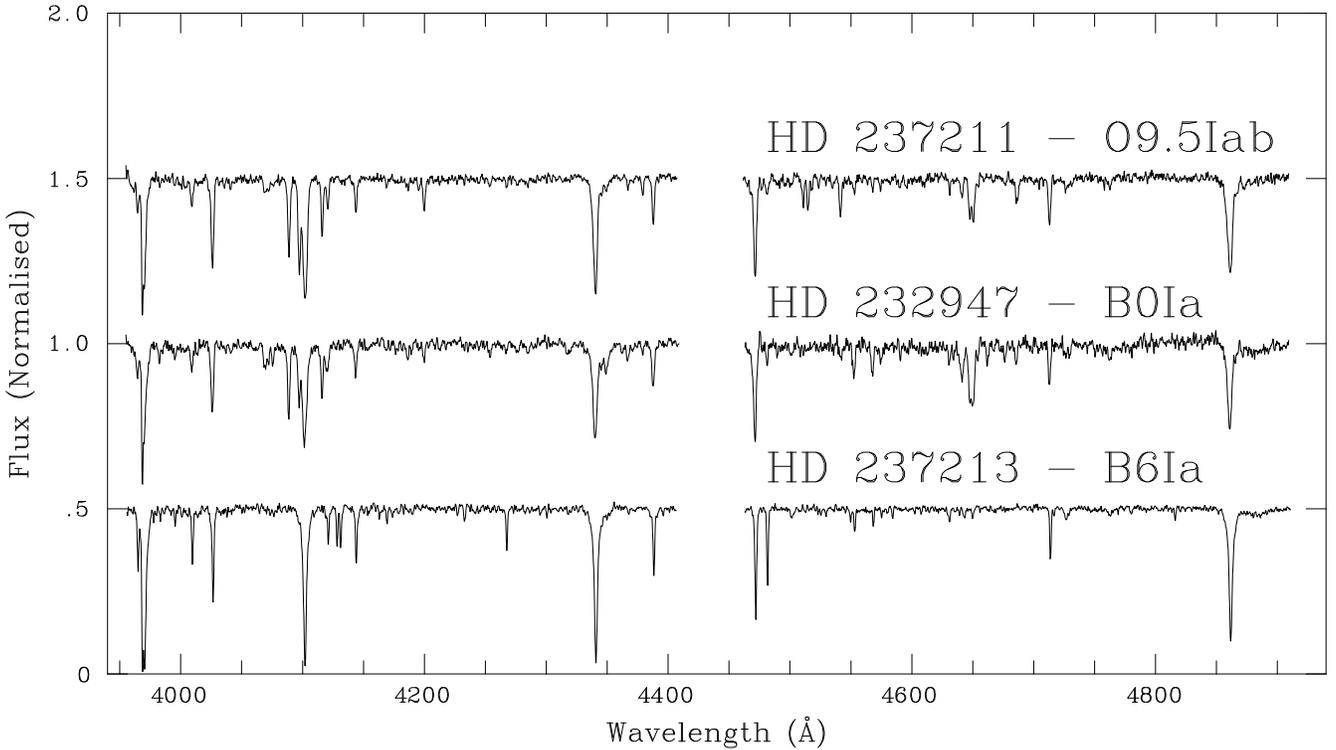}}
\end{picture}
  \caption{Blue spectra of supergiant stars in \object{Cam OB3}. The
   small gap around $\lambda = 4420$\AA\ indicates the division
   between the two poses. All these objects have $DM\geq 13.0$. }
   \label{fig:camob3}
\end{figure*}

We have obtained spectra for ten stars in the area of \object{Cam
OB3}, among which we note the following peculiarities:

\begin{itemize}

\item \object{HD 25914} is a known variable, \object{GQ Cam}, and radial
velocity measurements in the literature are rather inconsistent,
suggesting binarity. Its DM is likely then to be underestimated.

\item The spectral type for \object{HD 237213} is B6Ia, rather different
from the B3Ia given by \citet{mor55}. Photometric measurements by
\citet{fer83} and \citet{hil56} are rather discordant as well.

\item \object{BD +56\degr864} has been classified as O6Vnn. However, at the
resolution of our spectra, it is clear that the very broad lines are
due to the presence of two components in the lines. The two components
are much more clearly separated in the Position 2 spectrum than in
Position 1. \object{BD +56\degr864} is therefore a double-lined
spectroscopic binary. The two components have similar spectral types,
approximately O6V((f)), and therefore the DM is likely to be
underestimated by several tenths of a magnitude.
\end{itemize}

The eight objects with photoelectric photometry give an average
$\overline{DM}=13.0\pm0.2$. The low dispersion, specially when
considering that most stars observed are
supergiants, represents strong confirmation of the existence of
\object{Cam OB3}. \object{LS V $+56\degr$60}, with {\em Tycho}
photometry gives a larger distance modulus. If we transform the {\em Tycho}
magnitudes using Eq.~\ref{eq:ours} instead of Eq.~\ref{eq:tycho}, we
find $DM= 12.2$ and $13.3$ for \object{LS V $+56\degr$59} and
\object{LS V $+56\degr$60} respectively, compatible with
all the other stars in the area, if we take into account that
\object{LS V $+56\degr$59} is a double-lined spectroscopic binary.

The distance modulus to \object{Cam OB3} is almost two magnitudes
larger than would be expected if it was in the Perseus Arm. We
therefore conclude that \object{Cam OB3} is a tracer of the Outer
Arm. A more comprehensive study of this association will be presented
in future work.

\subsection{Auriga/Gemini} 
 
Galactic structure in the direction of the Anticentre is rather
complex. Traditionally, two OB associations have been considered to
exist in the range $l = 170\degr-180\degr$. In her
compilation of Galactic OB associations, \citet{humphreys} lists 12
likely members of \object{Aur OB1}, with an average distance modulus
$DM=10.6$, extending over a vast region spanning from $l_{\rm II} =
168\degr$ to $178\degr$ and from $b=-7.4\degr$ to
$b=+4.2$. Considering its distance, \object{Aur OB1} should be located in
the Orion Spur, but some authors \cite[e.g.,][]{mae95} doubt its
existence. 

 On the other hand, \object{Aur OB2} is given as
a rather more distant and compact association, extending between
$l_{\rm II} = 172\degr$ and $174\degr$ from $b=-1.8$ to $b=+2.0$.
The very young open cluster \object{NGC 1893}, containing several
O-type stars, has been traditionally considered its core. 
\citet{humphreys} lists only 8 members of \object{Aur OB2} and
adopts $DM=12.5$. This value, however, is just an average between the
distance to \object{NGC 1893} (taken as $DM=12.8$) and the rather
lower values for other presumed members. 

All modern work based on accurate photometry has resulted in much
larger distances to \object{NGC 1893}. \citet{fitz93} found $DM=13.4$
from Str\"{o}mgren photometry, while \citet{chopi} derived $DM=13.9$.
Such large values are incompatible with the distances to other
putative members and suggest that \object{Aur OB2} must be separated
into at least two groupings.

The two stars in our sample with Galactic longitude close to
$l=170\degr$  (\object{LS V $+40\degr$ 47} and \object{LS V $+33\degr$
36}) have $DM=13.6$ and $13.7$ respectively, in good agreement with the
distance to \object{NGC 1893}. In the interval $l=186\degr-192\degr$,
most stars in our sample have distances compatible with a prolongation
of the Perseus Arm, though the photometry is not very reliable for
some of our objects. Of particular interest is \object{LS V $+23\degr$
8}, which, if we trust the photographic photometry of \citet{mc67}, is
an obscured O8 supergiant lying in the Perseus Arm. This
possible prolongation of the Perseus Arm would be marked by objects in
\object{Aur OB2} not associated with \object{NGC 1893}, likely
connected with the young open clusters \object{Stock 8} \citep{mm71} and
\object{NGC 1931} \citep{pm86}, both at $DM\approx12$. Such distant
objects cannot be associated with the Orion Spur in this direction
($l=174\degr$).

Only three out of
nine objects
in this range have distances placing them close to the Cygnus Arm. 

\section{Discussion} 

\subsection{Validity of the method}
As outlined in Section~\ref{sec:method}, the method of spectroscopic
parallaxes cannot be considered very accurate in the determination of
the distance to a single star, but the different sources of
non-systematic uncertainties tend to cancel out when a large sample is
considered. Unfortunately, in this work we are covering a large span
of the Galactic plane and therefore the objects in any of the
individual areas surveyed represent only a moderately-sized sample.

In order to assess the dispersion that we could expect in our
determinations of $DM$, we observed with the OHP
1.93-m several pairs of stars of similar magnitude lying very close in
the sky (see Table~\ref{tab:ohp2}), in the hope that they might also
be physically related. In all cases, we found that the spectral types
of the two stars were very similar, suggesting that a physical
association did indeed exist. Of the seven cases considered, we find
five pairs of stars for which the difference in $DM$ is sufficiently
small to be explained solely by a dispersion of $\sim 0.3\:$mag in the
intrinsic magnitude calibration. The two exceptions, LS III
$+52\degr$17 \& 19 and LS III $+58\degr$70 \& 71, show differences
that would need a rather larger dispersion, but can also be explained
by assuming that one of the two stars is actually an unresolved binary.

\begin{table*}[ht]
\caption{Background early-type stars observed by \citet{mas95} in
Perseus Arm fields. Spectral types and photometric measurements are
from \citet{mas95}, while the corresponding $DM$ have been calculated
with the calibration used here.}
  \begin{center}
\begin{tabular}{lcccccc}
Field & Coordinates & $l$ & Spectral & $V$ &  $(B-V)$ & $DM$\\
& (J2000) & & Type &&\\
\hline
NGC 7235 & 22 13 30.34 +57 17 41.4 & $102\fdg8$&B1.5V & 14.67 & 0.98 & 13.8 \\
NGC 7380 & 22 47 30.93 +58 09 07.8 & $107\fdg2$&B3V & 13.89 & 0.44 & 13.6 \\
         & 22 47 52.01 +58 05 49.0 & $107\fdg2$&B0.5V & 15.53 & 1.12 & 15.1 \\
Cep OB5 & 23 01 44.67 +57 05 56.6 & $108\fdg4$ &B2V & 14.08 & 0.61 & 14.0 \\
        & 23 02 42.22 +56 57 14.0 & $108\fdg5$ &B1.5V & 14.48 & 0.75 & 14.3 \\
IC 1805 & 02 31 48.47 +61 34 55.8 & $134\fdg6$ &B2V & 13.93 & 0.76 & 13.4 \\
        & 02 32 06.48 +61 29 54.2 & $134\fdg6$ &B1.5V & 13.93 & 0.91 & 13.2 \\
        & 02 33 09.06 +61 27 46.1 &$134\fdg8$  &B1V & 12.93 & 0.65 & 13.4 \\
NGC 2244 & 06 31 37.10 +04 45 53.7 & $206\fdg4$ &B0.5V & 15.15 & 0.98 & 15.2 \\
         & 06 32 22.49 +04 55 34.4 & $206\fdg4$&B3V & 15.39 & 0.67 & 14.4 \\
         & 06 33 10.16 +04 59 50.2 & $206\fdg4$&B3V & 14.98 & 0.78 & 13.6 \\
\end{tabular}
\end{center}
 \label{tab:massey}
\end{table*}

As a matter of fact, the presence of unrecognised binaries in the
sample is likely to be a major contributor to the dispersion in $DM$
observed within one single area. In order to test the statistical
significance of the different groupings of stars, one would need to
observe a large sample of objects in a small region. Such a study will
be conducted in a forthcoming paper.

\subsection{Representativity of the sample}

The objective of this work was simply confirming the existence of OB
stars at distances compatible with the position of the Cygnus
Arm. Candidates were selected from the catalogues of luminous stars
because of their colours. Most of them are intrinsically very bright
objects with moderate reddening. As can be seen in
Tables~\ref{tab:ohp1} and~\ref{tab:ohp2}, most of the objects which
turn out to have $DM\geq13$ have luminosity classes I-III. This is a
bias inherent to the selection procedure, as we are picking up very
distant objects in a magnitude-limited sample.

Obviously, this sample is not representative of the population of OB
stars in the Cygnus Arm, but made up of the few bright objects that
happen to be seen through windows of low absorption in the Orion
Spur and the
Perseus Arm. A better idea of the sort of apparent magnitudes and
reddenings characterising the OB population of the Cygnus Arm can be
obtained with the following.

\begin{figure}
\begin{picture}(250,230)
\put(0,0){\includegraphics{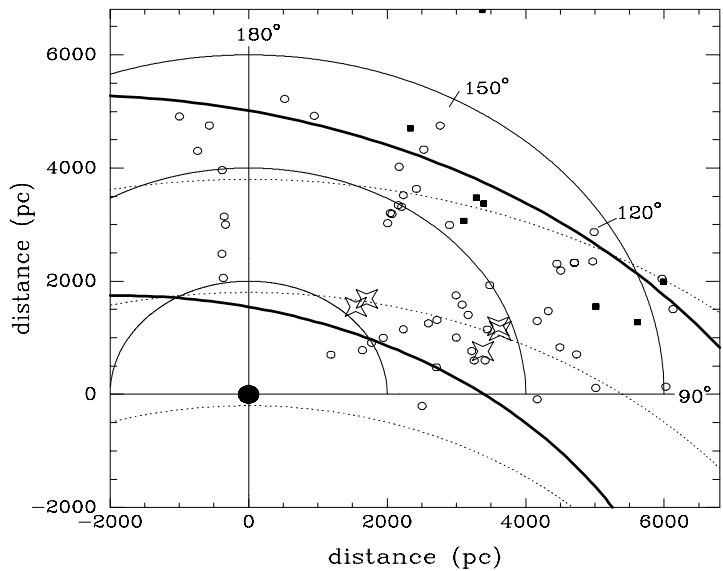}}
\end{picture}
  \caption{Position of the stars under investigation in the
   Galaxy. The black dot at the coordinate origin represents the
   position of the Sun. Dotted arcs represent circles of
   Galactocentric distance 7, 9 \& 11 kpc (assuming
   $R_{\sun}=7.2\:$kpc). The three semicircles correspond to distances of
   2, 4 and 6 kpc from the Sun, with Galactic longitude marked on the
   outermost semicircle. The thick traces are the logarithmic
   spiral arms corresponding to the Perseus Arm and Cygnus Arm in the
   model of \citet{val02}. Stars in our sample are shown as open
   circles, while stars from Table~\ref{tab:massey} are represented as
   filled squares. The large stars represent the locations of open
   clusters and associations considered to be good tracers of the Perseus Arm,
   \object{NGC 7235}, \object{NGC 7380}, \object{Cep OB5}, \object{IC
   1805} (distances from \citealt{mas95}) and \object{h Per} (distance
   from \citealt{mar01}).}
   \label{fig:mapa}
\end{figure}

In their comprehensive spectroscopic study of some OB associations of
the Northern Milky Way, \citet{mas95} obtained spectra of all stars
with blue colours in their fields, independently of magnitude. They
found a few OB stars far beyond 
the targeted fields in all the associations observed in the $l\approx
100\degr - 210\degr$ range, except in the case of \object{NGC 1893},
which is itself on the Cygnus Arm. All these objects are listed in
Table~\ref{tab:massey}, where we have calculated distance moduli for
them, based on the photometry of \citet{mas95}.

Five stars from Table~\ref{tab:massey} fall in the Cepheus
region. Among them, there is one with a very large $DM=15.1$. The
other four average to $\overline{DM}=14.0\pm0.3$, in good agreement
with our sample. Likewise, the three objects beyond the field of
\object{IC 1805} ($l\approx136.6\degr$) agree quite well on
$\overline{DM}=13.3$, quite similar to the values found for our
objects at $l\approx120\degr$ and $l\approx147\degr$.

The objects in Table~\ref{tab:massey} have on average spectral types
indicating that they are rather less luminous than stars in our
sample. They are also rather more obscured, which is consistent with
their location behind star-forming regions of the Perseus Arm.
As a consequence, they are several magnitudes fainter than objects in
our sample. However, they are all within the reach of a moderate-size
telescope with a modern spectrograph. Hence, it is to be expected that
a sizable proportion of the OB population of the Cygnus Arm may be
identified.

\subsection{Is the Cygnus Arm well traced?}

\citet{mof79} estimate a $DM=13.2\pm0.2$ for the young open cluster
\object{Waterloo 1}, at $l=151.4\degr$. For two other small groups of
young stars associated with the \ion{H}{ii} regions \object{Sh 2-217}
and \object{Sh 2-219} (at $l\approx159.3\degr$), both of which contain
embedded clusters \citep{deh03}, they estimate
$DM=13.1\pm0.3$ and $DM=13.6\pm0.3$, while for \object{NGC 1624} (at
$l=155.4\degr$), they estimate $DM=13.9\pm0.2$. All these estimates
are in good agreement with the $DM$ we find for the nearby \object{Cam
OB3}. 

Three stars close to $l=135\degr$ in Table~\ref{tab:massey} yield
$\overline{DM}=13.3$, intermediate between the values obtained for
presumed tracers of the Cygnus Arm in the Cassiopeia region
(Section~\ref{sec:cas}) and for the \object{Cam
OB3} region. All these values support the idea
of a coherent spiral structure with $d=4-5$ kpc over the
$l=130\degr-150\degr$ range. This distance is slightly shorter than
predicted by the models of \citet{val02} or \citet{rus03} (in which the
Cygnus Arm runs at $\approx 5$ kpc), but certainly consistent with them,
if we take into account the width of a spiral arm.
  
This structure is likely to be continued by the objects we find at
$d=5.3$ kpc around $l=120\degr$ and the objects at $d=6-7$ kpc at
$l\approx 105\degr$. However, a larger sample of distant stars and
some more secure distances (such as good photometric distances to open
clusters) will be necessary in order to confirm the continuity of the
Cygnus Arm over such a large span.

Over the $l=160\degr-170\degr$ range there is a lack of possible Outer
Arm tracers in the literature, but beyond $l=170\degr$ there is a
number of open clusters clearly delineating it (\object{NGC 1893},
\object{Bochum 1} and 4 other clusters studied by \citealt{tm93}).
The existence of
a structure at $d=5-6$ kpc over the $l=175\degr-215\degr$ range seems
rather secure.

\section{Conclusions}
We have carried out a spectroscopic survey of stars in the $l=85\degr
- 190\degr$ range reported in the literature to have photometric
indices suggesting that they could be very
distant OB stars. Over the whole range, we find a large number of
objects whose spectroscopic parallaxes support a distance well above
that expected for Perseus Arm stars and in relatively good agreement
with the predictions for the position of the Outer or Cygnus Arm.

 Over the $l=85\degr - 115\degr$ range, tracers of the Cygnus Arm
are not very numerous and sparsely located. Many of them are stars of
high luminosity, suggesting that at these Galactic longitudes the main
sequence B-star population of the Cygnus Arm is far too faint to appear
in the LS catalogues even in the regions of lower extinction.

 From $l=115\degr$, tracers of the Cygnus Arm start to be more
frequent. Several objects in the field of \object{Cas OB5} and
\object{Cas OB4} give distance moduli clustered around $DM=13.6$,
suggesting that the Cygnus Arm is there at $d\approx5.3\:$kpc.

 In the $l=140\degr-180\degr$ range, Perseus Arm tracers are rather
inconspicuous, with the possible exception of some objects in \object{Aur
OB2}. Over this range, most spiral tracers seem to belong to the Outer
Arm, including the extended \object{Cam OB3} association at $d \approx
4\:$kpc and \object{NGC 1893} and its surrounding area, which should
be separated from other clusters in \object{Aur OB2}.

\begin{acknowledgements} 

IN would like to thank all the staff at the Observatoire de
Haute Provence for their support and friendliness and Dr. GianLuca
Israel for help during the NOT observations.

Part of the data presented here have been taken using ALFOSC, which is 
owned by the Instituto de Astrof\'{\i}sica de Andaluc\'{\i}a (IAA) and
operated at the Nordic Optical Telescope under agreement
between IAA and the NBIfAFG of the Astronomical Observatory of
Copenhagen. This research has made use of the Simbad data base,
operated at CDS, Strasbourg, France.   

We thank an anonymous referee for comments that improved the
presentation and readability of the paper.

During part of this research, IN has been supported by a contract
under the programme ``Ram\'on y Cajal'' of the Spanish Ministerio de
Ciencia y Tecnolog\'{\i}a.
This research is partially supported by the Spanish Ministerio de
Ciencia y Tecnolog\'{\i}a under grant AYA2002-00814.

\end{acknowledgements}


\begin{thebibliography}{}

\bibitem[Baade(1983)]{baa83} Baade, D. 1983, A\&AS, 51, 235

\bibitem[Bianchi \& Garc\'{\i}a(2002)]{bg02}Bianchi, L., \& Garc\'{\i}a,
M. 2002, ApJ 581, 610

\bibitem[Chavarr\'{\i}a-K. et al.(1994)]{cha94}Chavarr\'{\i}a-K., C.,
Moreno-Corral, M.A., Hern\'{an}dez-Toledo, H., et al. 1994, A\&A 283, 963

\bibitem[Dame et al.(2001)]{dam01} Dame, T.M., Hartmann, D., \&
  Thaddeus, P. 2001, ApJ, 547, 792 

\bibitem[Deharveng et al.(2003)]{deh03} Deharveng, L., Zavagno, A.,
Salas, L., et al. 2003, A\&A 399, 1135

\bibitem[Draper et al.(2000)]{draper} Draper, P.W., Taylor, M. , \&
  Allan, A. 2000, Starlink User Note 139.12, R.A.L. 

\bibitem[Drilling(1975)]{dri75} Drilling, J.S. 1975, AJ, 80, 128

\bibitem[Drimmel \& Spergel(2001)]{das01}Drimmel, R., \& Spergel,
  D.N. 2001, ApJ 556, 181 

\bibitem[Fern\'{a}ndez et al.(2001)]{fer01}Fern\'{a}ndez, D.,
Figueras, F., \& Torra, J. 2001, A\&A 372, 833 

\bibitem[Fernie(1983)]{fer83} Fernie, J.D. 1983, ApJS, 52, 7 

\bibitem[Fitzpatrick(1999)]{fit99} Fitzpatrick E.L., 1999, PASP 111, 
63 

\bibitem[Fitzsimmons(1993)]{fitz93} Fitzsimmons, A., 1993, A\&AS, 99,
15 

\bibitem[Forbes(1985)]{for85} Forbes, D., 1985, AJ, 90, 301 

\bibitem[Garmany \& Stencel(1992)]{gs92} Garmany, C.D., \& Stencel,
R.E. 1992, A\&AS 94, 211 

\bibitem[Georgelin \& Georgelin(1976)]{gag}Georgelin, Y.M., \& 
 Georgelin, Y.P. 1976, A\&A, 49, 57 

\bibitem[Gillet et al.(1994)]{gil94}Gillet, D., Burnage, R., Kohler, D., et
  al. 1994, A\&AS, 108, 181 

\bibitem[Haug(1970)]{haug} Haug., U. 1970, A\&AS, 1, 35

\bibitem[Herrero et al.(2000)]{her00} Herrero, A., Puls, J., \&
Villamariz, M. R. 2000, A\&A, 354, 193 

\bibitem[Hiltner(1956)]{hil56}Hiltner, W.A. 1956, ApJS, 2, 389 

\bibitem[H\o g et al.(2000)]{tycho} H\o g, E., Fabricius, C., Makarov,
V.V., et al. 2000, A\&A, 355, L27  

\bibitem[Howarth et al.(1997)]{howarth}Howarth, I., Murray, J.,  
Mills, D., \& Berry, D.S. 1997, Starlink User Note 50.20, R.A.L. 

\bibitem[Humphreys(1978)]{humphreys} Humphreys, R.M. 1978, ApJS 38, 
309  

\bibitem[Humphreys \& McElroy(1984)]{hme84} Humphreys, R.M., \& McElroy, D.B. 1984, ApJ 284, 565 

\bibitem[Isserstedt(1970)]{iss70} Isserstedt, J. 1970, A\&A, 9, 70 

\bibitem[Kimeswenger \& Weinberger(1989)]{kw89} Kimeswenger., S, \&
  Weinberger, R. 1989, A\&A, 209, 51 

\bibitem[Kulkarni et al.(1982)]{kul82} Kulkarni, S.R., Blitz, L., \&
Heiles, C., 1982, ApJ, 259, L63

\bibitem[Lennon et al.(1990)]{len90} Lennon, D.J., Dufton, P.L.,
    Fitzsimmons, A., et al. 1990, A\&A, 240, 349 

\bibitem[Lennon et al.(1992)]{len92} Lennon, D.J., Dufton, P.L.,
    \& Fitzsimmons, A. 1992, A\&AS, 94, 569 

\bibitem[Marco \& Bernabeu(2001)]{mar01} Marco, A., Bernabeu, G. 2001,
A\&A, 372, 477 

\bibitem[Marco et al.(2001)]{chopi} Marco, A., Bernabeu, G., \&
  Negueruela, I. 2001, AJ, 121, 2075  

\bibitem[Martin(1972)]{mar72} Martin, N. 1972, A\&A 17, 253

\bibitem[Martins et al.(2002)]{mar02}Martins, F., Schaerer, D., \&
Hillier, J. 2002, A\&A, 382, 999  

\bibitem[Massey et al.(1995)]{mas95} Massey, P., Johnson, K.E., and
  DeGioia-Eastwood, K. 1995, ApJ, 454, 151

\bibitem[Mathys(1988)]{mat} Mathys, G. 1988, A\&AS 76, 427

\bibitem[Mayer \& Macak(1971)]{mm71} Mayer, P., \& Macak, P. 1971, 
Bull. Astron. Inst. Czech., 22, 46  

\bibitem[McCuskey(1967)]{mc67} McCuskey, S.W. 1967, AJ 72, 1199

\bibitem[Mel'nik \& Efremov(1995)]{mae95} Mel'nik, A.M., \& Efremov,
Yu.N. 1995, AstL 21, 10

\bibitem[Moffat(1971)]{mof71} Moffat, A.F.J. 1971, A\&A 13, 20

\bibitem[Moffat \& Vogt(1973)]{mav73} Moffat, A.F.J., \& Vogt, 
  N. 1973, A\&A, 30, 381 

\bibitem[Moffat \& Vogt(1975)]{mav75} Moffat, A.F.J., \& Vogt, 
  N. 1975, A\&AS, 20, 85 

\bibitem[Moffat et al.(1979)]{mof79} Moffat, A.F.J., Fitzgerald, M.P.,
\& Jackson, P.D. 1979, A\&AS 38, 197

\bibitem[Morgan et al.(1952)]{mor52} Morgan, W.W., Sharpless, S., \& 
  Osterbrock, D. 1952, AJ, 57, 3 

\bibitem[Morgan et al.(1955)]{mor55} Morgan, W.W., Code, A.D., \& 
Whitford, A.E. 1953, ApJS, 2, 41 
 
\bibitem[Munari \& Carraro(1995)]{mac95} Munari, U., \& Carraro, 
  G. 1995, MNRAS, 277, 1269 

\bibitem[Pandey \& Mahra(1986)]{pm86}Pandey, A.K., \& Mahra,
H.S. 1986, Ap\&SS, 120, 107

\bibitem[Quillen(2002)]{qui02} Quillen, A.C. 2002, AJ 124, 924

\bibitem[Russeil(2003)]{rus03} Russeil, D. 2003, A\&A 397, 133

\bibitem[S\"{a}rg \& Wramdemark(1977)]{sw77}S\"{a}rg, K., \&
Wramdemark, S. 1977, A\&AS 27, 403

\bibitem[Shortridge et al.(1997)]{shortridge}Shortridge, K.,
Meyerdicks, H., Currie, M., et al. 1997, Starlink User Note 86.15,
R.A.L 

\bibitem[Turbide \& Moffat(1993)]{tm93}Turbide, L., \& Moffat,
A.F.J. 1993, AJ 105, 1831

\bibitem[Turner(1980)]{tur80} Turner, D.G. 1980, ApJ, 235, 146

\bibitem[Vacca et al.(1996)]{vac96} Vacca, W.D., Garmany, C.D., \&
 Shull, J.M. 1996, ApJ 460, 914

\bibitem[Vall\'{e}e(2002)]{val02} Valle\'{e}e, J.P. 2002, ApJ,
 566, 261

\bibitem[Wakker \& van Woerden(1991)]{ww91}Wakker, B.P., \& 
 van Woerden, H. 1991, A\&A, 250, 509

\bibitem[Walborn(2002)]{wal02}Walborn, N.R. 2002, AJ, 124, 507

\bibitem[Walborn \& Fitzpatrick(1990)]{waf}Walborn, N.R., \& 
 Fitzpatrick, E.L. 1990, PASP, 102, 379

\bibitem[Wegner(1994)]{weg94} Wegner, W. 1994, MNRAS, 270, 229

\bibitem[Wramdemark(1976)]{wram76} Wramdemark, S. 1976, A\&AS, 26, 31  

\bibitem[Wramdemark(1981)]{wram81} Wramdemark, S. 1981, A\&AS, 43, 103
 
\end{thebibliography}
\end{document}